# Assessing Response Disparities in California Wildland-Urban-Interface (WUI) Cities Using the Compartmental Model

Zihui Ma[1], Guangxiao Hu[2*], Ting-Syuan Lin[3], Lingyao Li[4], Songhua Hu[5], Loni Hagen[4], and Gregory B. Baecher[1]

[1] Department of Civil and Environmental Engineering, University of Maryland, MD, United States
[2] Center for Global Sustainability, School of Public Policy, University of Maryland, MD, United States
[3] Department of Sociology, University of Illinois at Urbana-Champaign, IL, United States
[4] School of Information, University of South Florida, FL, United States
[5] Senseable City Lab, Massachusetts Institute of Technology, MA, United States
[*] corresponding author

**Abstract:**

The increasing frequency and severity of wildfires pose significant risks to communities, infrastructure, and the environment, especially in Wildland-Urban Interface (WUI) areas. Effective disaster management requires understanding how the public perceives and responds to wildfire threats in real-time. This study uses social media data to assess public responses and explores how these responses are linked to city-level community characteristics. Specifically, we leveraged a transformer-based topic modeling technique called BERTopic to identify wildfire response-related topics and then utilized the Susceptible-Infectious-Recovered (SIR) model to compute two key metrics associated with wildfire responses — awareness and resilience indicators. Additionally, we used GIS-based spatial analysis to map wildfire locations along with four groups of city-level factors (racial/ethnic, socioeconomic, demographic, and wildfire-specific). Our findings reveal significant geographic and socio-spatial differences in public responses. Southern California cities with larger Hispanic populations demonstrate higher wildfire awareness and resilience. In contrast, urbanized regions in Central and Northern California exhibit lower awareness levels. Furthermore, resilience is negatively correlated with unemployment rates, particularly in southern regions where higher unemployment aligns with reduced resilience. These findings highlight the need for targeted and equitable wildfire management strategies to improve the adaptive capacity of WUI communities.

**Keywords:** Wildfire Response; Wildland-Urban Interface (WUI) communities; Social Inequity; Equitable strategy; Social Media Analysis

**Significance Statement**

Our study used Census and Twitter data to estimate levels of wildfire responses in impacted areas through the integrations of BERTopic and SIR modeling. These quantifiable metrics can assist governments in making timely and equitable decisions based on social media conversations during disasters. More specifically, the visualization, mixed with the awareness and resilience measures, can be adopted to the crisis management dashboard. In our case analysis, we discovered disparities in the level of awareness and resilience, informing culturally sensitive, regionally tailored, and demographically targeted approaches in wildfire awareness campaigns and emergency management. Our findings further highlight the importance of future collaboration with local government, regional planners, and crisis management scholars to explore structural inequality and inform policy updates.

**Introduction and Background**

Climate change has led to rising global temperatures, prolonged droughts, and reduced rainfall, significantly increasing the frequency and intensity of wildfires worldwide (Burke et al. 2021; Synolakis and Karagiannis 2024). In California, for example, the fire season now spans most of the year due to record-high temperatures and persistent drought, with the state experiencing some of the largest wildfires in its history in the past few decades. A recent study published in *Nature* (Brown et al. 2023) concluded that anthropogenic warming has enhanced the aggregate expected frequency of extreme daily wildfire growth in California by an average of 25% relative to preindustrial conditions. As climate change continues to alter ecosystems and weather patterns, studies have predicted that the frequency of wildfires will continue to rise, threatening both human communities and natural environments worldwide (Brown et al. 2023; Di Virgilio et al. 2019).

Wildland-urban interface (WUI) areas, where human developments meet with wildland vegetation, have expanded significantly due to population growth and urban sprawl (Fu et al. 2023). These areas are particularly vulnerable to wildfires, experiencing devastating economic losses, displacement of residents, and long-term ecological damage (Holder et al. 2023). According to the U.S. Department of Agriculture Forest Service (Mockrin 2024), between 1990 and 2020, the size of the WUI increased by 179,000 square kilometers, and the number of houses in the WUI grew by 46%, from 30 million to 44 million. This expansion has dramatically increased the potential for wildfire disasters and complicated emergency response efforts, as more people and properties are now situated in fire-prone areas. Therefore, it is critical to develop more efficient, cost-effective, and time-sensitive methods to monitor how communities in WUI areas respond during wildfires.

The expansion of WUI areas has also intensified socio-spatial disparities in human responses to wildfires. Previous studies (Davies et al. 2018; Meldrum et al. 2018; Yadav et al. 2023) have documented that socioeconomic and demographic factors such as income levels, education attainment, ethnicity, and age distribution are significantly associated with vulnerability and resilience during wildfires. For example, lower-income households often lack the financial means to implement effective fire mitigation strategies (Masri et al. 2021); language barriers can impede access to critical emergency information for non-English-speaking populations, delaying their response time (Yadav et al. 2023). Socioeconomic status can also affect access to information and perception of risk: communities without reliable internet access may not receive timely updates on wildfires (Fowler et al. 2019), and education levels can influence how individuals perceive risks and adhere to evacuation orders (Oliveira et al. 2020). Understanding these unique challenges across different communities is essential for shaping emergency management strategies that promote more equitable preparedness, evacuation procedures, and recovery efforts.

In recent decades, significant efforts have been made in wildfire monitoring to reduce potential risks. Traditional approaches have primarily focused on the physical and environmental impacts of wildfires using meteorological data and combustible material information (Hanes et al. 2023; Júnior et al. 2022). While these methods have been instrumental in predicting and estimating the direct impact and spread of wildfires, they are typically time-insensitive and unable to capture the dynamic, real-time public perceptions and responses, which are critical during emergencies for informing optimal evacuation policies. The rise of social media has added new dimensions to public engagement and information dissemination during disasters (Imran et al. 2020; Ma, Li, Mao, et al. 2024). Platforms such as Twitter (now X), Facebook, and Instagram have become essential tools for timely updates, resource sharing, evacuation, and crowdsourced crisis mapping. However, despite the recognized importance of social media in disaster response, research has largely focused on its application in hurricanes and earthquakes (Li, Ma, et al. 2023; Li, Bensi, and Baecher 2023; Mirbabaie et al. 2020), with comparatively limited attention to wildfires (Li, Ma, and Cao 2021; Yue et al. 2021). In addition, existing studies (Gong, Dimitrov, and Bartolacci 2024; Michail et al. 2024) mainly focus on wildfire detections, evacuation measures, and smoke dispersion through network and textual analysis of social media content, but none have quantified how individual perceptions and community awareness dynamically evolve during wildfires. Furthermore, while socio-spatial disparities in wildfire impacts have been well-documented (Davies et al. 2018; Meldrum et al. 2018; Yadav et al. 2023), most research focuses on disparities in wildfire preparedness, damage, and recovery, while there remains a gap in examining the disparities in public responses during wildfires and how these responses evolve dynamically across different communities.

To address these gaps, more comprehensive research is needed that analyzes not only the content of social media posts during wildfires but also examines the dynamics of public awareness and resilience across communities with different socio-spatial characteristics. Accordingly, two key research questions (RQs) are proposed:

RQ(1): How can social media data be leveraged to develop numerical metrics that capture communities' awareness and engagement in wildfire responses? How are these metrics geographically distributed?

RQ(2): What are the relationships between city-level characteristics (e.g., socioeconomic, demographic, and wildfire intensity) and social responses? How do the characteristics of different communities affect their responses to wildfires at the city level?

Specifically, we collected data from Twitter during the 2020 California wildfires, employing natural language processing (NLP) and topic modeling techniques to extract information relevant to public responses to wildfires. By integrating the identified topics with compartmental epidemiological models, the Susceptible-Infected-Recovered (SIR) model, we estimated public response for each city from two perspectives: awareness—the degree to which communities are informed regarding wildfire threats; and resilience—the capacity of communities to recover after wildfires. Finally, we examined city-level socio-spatial disparities via a Multiscale Geographically Weighted Regression (MGWR), accounting for regional variations in relationships across factors such as income levels, educational attainment, ethnicity, age distribution, and historical wildfire risk. Our study provides a faster and more comprehensive way for mapping wildfire response assessments, enabling governments to quickly gauge public resilience and adaptability during wildfires. In addition, the documented socio-spatial disparities, with their spatial variations, provide valuable insights for developing more targeted policy recommendations to build a more equitable and resilient society.

*Social media in wildfire response applications*

With the increasing frequency and severity of wildfires in the expanding WUI, it has become essential to understand public reactions to enhance decision-making. Traditionally, wildfire management strategies focused on reducing vulnerability to protect people, property, and ecosystems (Thompson et al. 2019; Tymstra et al. 2020). Systems like the Canadian Forest Fire Danger Rating System (Hanes et al. 2023) and the U.S. National Fire Danger Rating System (Júnior et al. 2022) typically rely on meteorological data and combustible materials for risk assessment but tend to overlook key elements like topography and human activities. Recently, advancements in satellite remote sensing have improved data acquisition and processing, allowing the integration of diverse wildfire drivers such as vegetation and weather conditions (Thangavel et al. 2023; Yang, Jin, and Zhou 2021; Zhu et al. 2023). Researchers have developed models that consider the complex interactions among these factors using advanced artificial intelligence techniques like neural networks and support vector machines to enhance predictive accuracy and efficiency. For instance, Shao et al. (2023) developed a wildfire risk prediction model using network models to analyze the spatial relationships of active fire hotspots. Jiang et al. (2024) employed a deep learning-based Convolutional Neural Network (CNN) model that extracts high-level features from various wildfire-driven factors, facilitating accurate wildfire occurrence predictions. However, despite their insights into the physical and environmental aspects of wildfires, these methods often overlook the real-time responses of communities affected by such emergencies.

Social media enhances disaster management by offering a rapid, widespread platform for real-time information dissemination during emergencies, especially when traditional communication methods are unavailable or data collection after the disaster is resource-intensive (Imran et al. 2020; Ma, Li, Mao, et al. 2024). Its immediacy and broad reach greatly improve various aspects of disaster response, from monitoring the evolution of public concerns (Cai, Luo, and Cui 2021; Chen and Sun 2016; Mihunov et al. 2022; Rachel J, A, and M 2024; Wahid et al. 2022) to conducting emotional and behavioral analyses (García et al. 2024; Li et al. 2024; Li, Ma, and Cao 2020; Momin, Hasnine, and Sadri 2024; X. Wang et al. 2024). For example, C. Wang et al. (2024) utilized the Latent Dirichlet Allocation (LDA) model to explore topics related to public concerns during the 2021 earthquake in Qinghai Province, analyzing their evolution over time. Their analysis identified common themes in online public opinion and suggested that variations in the popularity of these opinions were linked to aftershock occurrences. Similarly, Yuan, Li, and Liu (2020) applied the LDA topic model to examine public expression throughout different disaster periods. This study introduced innovative techniques such as weights and weighted sentiment to quantify and analyze these topics, providing insights into public response trends during Hurricane Matthew and offering strategic guidelines for future responses. However, these methods focused on broad overviews and lacked insights into how different cities or regions specifically responded to disasters. Despite advancements in machine learning, there also remains a gap in utilizing advanced natural language processing tools to efficiently explore city-level responses.

Additionally, while current research has focused more on hurricane and earthquake events, the application of social media in WUI community responses remains underdeveloped. Research has primarily focused on wildfire detections (Gong et al. 2024; Michail et al. 2024), evacuation measures (Hocevar 2024; Li et al. 2021), and smoke dispersion modeling (Slavik et al. 2024; VanderMolen et al. 2024), which exposes a significant gap in understanding of dynamic community reactions. Traditional analytical methods also lack the robust quantitative tools that decision-makers need to effectively compare impacts across different regions. This underscores more sophisticated analytical tools that can transform social media data into actionable intelligence for diverse community scenarios, thereby enabling more strategic and informed decision-making in wildlife disaster management.

*Assessment of social disparities of disaster responses*

As climate change intensifies the frequency and severity of natural disasters, it disproportionately affects different communities, highlighting the need to address disparities in disaster management. Socio-demographic factors, such as income levels, race, and age, influence communities' experience and cope with disasters. Identifying vulnerable subgroups and ensuring that they receive appropriate support is crucial for improving resilience and recovery efforts (Cutter and Finch 2008; Doorn 2017). For example, Smiley et al. (2022) investigated the social inequalities in climate change-attributed extreme weather events, indicating that climate justice challenges affected vulnerable populations disproportionately in Latina/x/o neighborhoods. Similarly, (Cocina Díaz, Llorente-Marrón, and Dema Moreno 2024) found that female-headed households were more severely affected by earthquakes. Other studies have shown that racial minorities, children, and the elderly are more vulnerable due to factors such as substandard housing, limited insurance, and restricted access to education and transportation (Flanagan et al. 2011; Ranabir, Eghdami, and Singh 2014; Sanders et al. 2023; Tate et al. 2021; Wing et al. 2022).

Geographic factors also play a critical role in disaster impacts. Spatial analyses have highlighted the uneven distribution of disaster risk exposure, often correlated with socio-demographic characteristics (Chakraborty et al. 2022, 2023; Mazumder, Landry, and Alsharif 2022). For example, Chakraborty et al. (2023) identified hotspots of social vulnerability and flood exposure in urban areas, advocating for policy interventions to safeguard at-risk populations. Likewise, (Mazumder et al. 2022) emphasized spatial disparities in flood exposure based on age, income, and education, highlighting the necessity for targeted policies. However, most of these studies have focused on direct effects, which do not fully capture the adaptive capacities and real-time needs of different communities. The role of individual perceptions and community awareness in shaping resilience is often overlooked. Additionally, individuals and communities actively modify their levels of exposure and vulnerability, contributing to dynamic responses that are rarely considered in social inequity assessments aimed at enhancing resilience.

On the other hand, communities in WUI areas are increasingly vulnerable and have become a growing concern due to the significant risks of property damage and environmental and social disruption (Greenberg et al. 2024; Thomas et al. 2022). Ager et al. (2019) demonstrated that many communities in the western U.S. face heightened wildfire exposure due to their proximity to national forests. Similarly, Yadav et al. (2023) found that disadvantaged communities in

California, USA, are increasingly affected by wildfires, underscoring the need for equitable wildfire management strategies. Although research (Davies et al. 2018; Lambrou et al. 2023)(Davies et al. 2018; Lambrou et al., 2023) has explored how social and environmental vulnerabilities influence wildfire preparedness and recovery, there remains a pressing need to examine spatial and socio-demographic disparities in public responses to wildfires. This is particularly important given the geographic and social diversity within WUI areas (Meldrum et al. 2018). Future research needs to delve deeper into these disparities to inform more targeted and equitable wildfire management strategies.

**Study Areas**

The study area is the entire state of California (CA), known for its diverse landscapes and extensive WUI regions, where human habitation intersects with undeveloped wildland, increasing wildfire risks (Jones 2024; Mockrin et al. 2022). Recent study using remote sensing data reveals that approximately 6.74% of California's total land area, or about 423,971 square kilometers, falls within the WUI (Li et al. 2022). This area is primarily along the western coastline and west of the Sierra Nevada Mountain range, with lower density in the central and southeastern parts of the state. Urban sprawl, climate change, and lifestyle choices are expanding the WUI, placing more communities at risk (Ager et al. 2019; Radeloff et al. 2018; Xu et al. 2022).

Our case study examines the 2020 California wildfire season, the largest on record, which burned over 4 million acres (CAL FIRE 2021; Stelloh 2020). Notably, the August Complex Fire became the first "gigafire" in modern history, burning over 1 million acres (Milman and Ho 2020). Other significant fires included the Creek Fire, which trapped hundreds of campers near the Mammoth Pool Reservoir, and the North Complex, which caused widespread evacuations and 16 fatalities as it approached Oroville (ABC 7 news 2020). These fires, fueled by dry vegetation, high winds, and drought, severely impacted air quality, displaced thousands, and highlighted the urgent need for improved forest management and public preparedness, especially in WUI areas.

**Results**

In response to RQ (1), we display the distribution of two response indicators in Figure 2. Within the present social media landscape, the data available to infer the parameters of a SIR model are sometimes insufficient, discretized to unit time intervals, or lack records of changes in the size of the susceptible compartment. This probability is due to the dynamics of social media usage and digital coverage variability. Bernabeu-Bautista et al. (2021) demonstrated that urban areas exhibit higher online activity compared to mountain regions, while Sapienza et al. (2023) further documented diverse social media behaviors across various geographic areas. Therefore, to effectively model the SIR models, we filtered out cities where tweets did not occur continuously or where only a few tweets were posted per day. As a result, we included 109 cities in our SIR models, with their awareness indicators shown in Figure 1(a) and resilience indicators in Figure 1(b). The map also features the official fire perimeters, outlined in dashed red lines for reference. For a more detailed analysis, we focused on three major WUI areas near San Francisco (tag A), Fresno (tag B), and Los Angeles (tag C) counties.

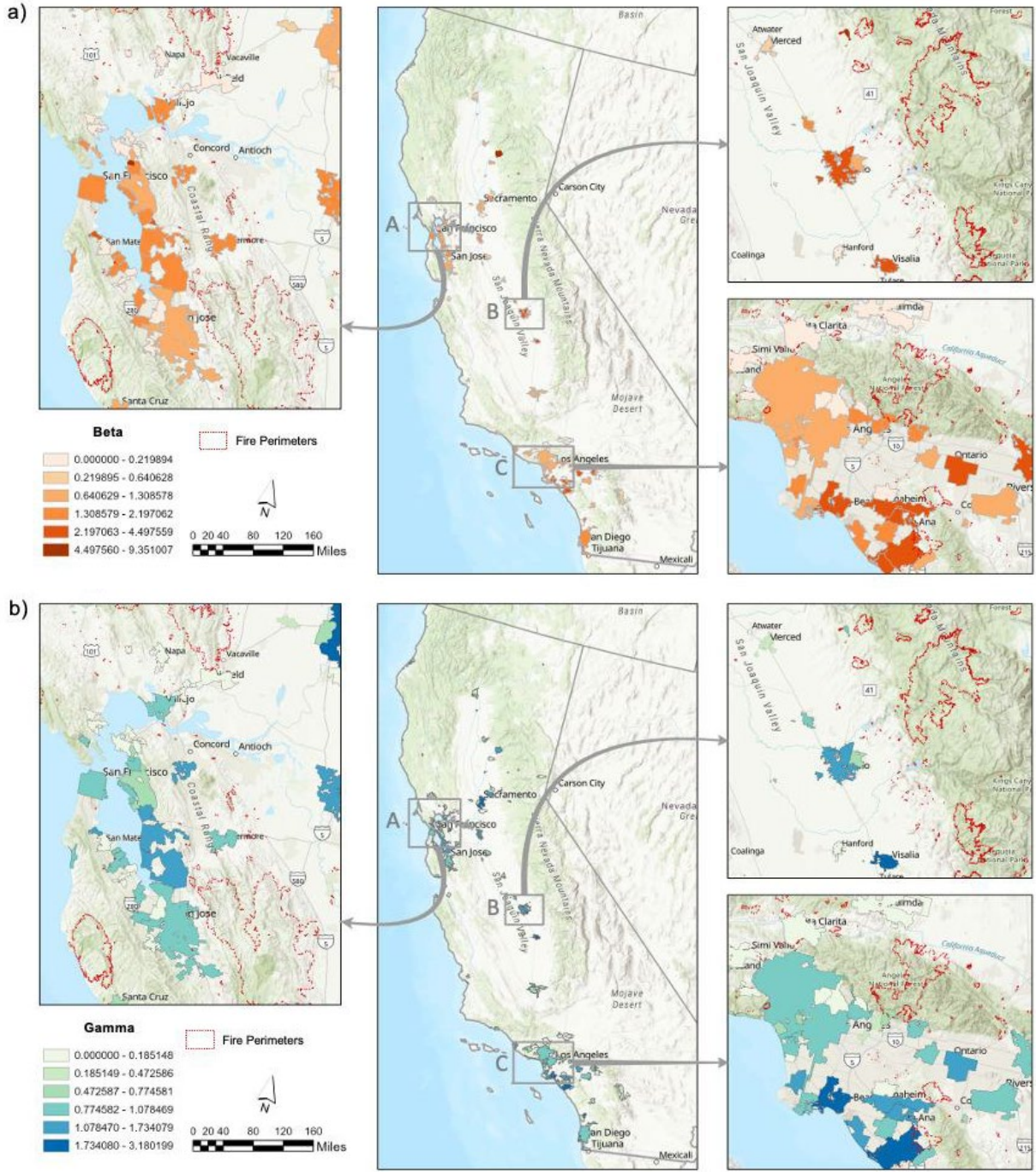

Figure 1. The overall public response distribution map of a) awareness indicators; b) resilience indicators.

We observed that two indicators closely mirrored the boundaries of wildfires, showing increased online engagement near fire zones. Specifically, the North Complex Fire (Rosoff 2020) —the deadliest of the season—and other major fires such as the SCU, LNU, and CZU Lightning Complex fires, significantly impacted Northern California (Bloch, Bogel-Burroughs, and Rio 2020), particularly the San Francisco area (tag A). Additionally, the Creek Fire, a major event in Central California's Sierra National Forest (Staff 2020), profoundly impacted the Fresno area (tag

B), leading to heightened awareness and response efforts. In contrast, Southern California, including Los Angeles County, experienced fewer but substantial fires like the Bobcat Fire, which was among the largest in the county's history and led to significant structural damage and evacuations (Member 2020; NASA 2020).

Figure 1 shows that the study areas displayed strong awareness and resilience indicators, particularly near-fire perimeters, indicating an effective community response to wildfires. Notably, on September 9th, San Francisco Bay Area (tag A) residents experienced unusually dark skies due to smoke from the North Complex Fire (including the Bear Fire) (Hartlaub 2020), prompting widespread public alertness and media coverage. The Creek Fire's rapid expansion from September 6-9 was fueled by strong winds and a significant buildup of dead trees (Yurong 2020), drawing significant attention on social media and prompting evacuation orders and road closures by authorities. These incidents highlight the communities' heightened awareness and swift responses, which enhanced resilience as depicted in our mapping analysis.

However, certain cities like Altadena issued evacuation warnings as the Creek Fire approached neighboring areas such as Arcadia and Pasadena. These areas demonstrated a lower level of public awareness, as reflected by their subdued response levels (Figure 1(a)). Additionally, the CZU Lightning Complex fire impacted regions in San Mateo and Santa Clara Counties, where cities like San Mateo, Menlo Park, and Stanford also showed lower response levels compared to more densely populated areas in the county such as Redwood City and Cupertino (Figure 1(a)). This disparity in response levels across different cities underscores the need for tailored response strategies that address each community's specific awareness and preparedness needs.

Moreover, our observations suggest that locations with high awareness are often consistent with areas of high resilience, indicating that communities most aware of wildfire threats tend to exhibit stronger responses and greater resilience to such events. For instance, even though Los Angeles (tag C) faced fewer fires, the twitter awareness raised during the Bobcat Fire helped facilitate a well-coordinated community response, illustrating how informed communities are better prepared to handle such crises. As depicted in Figure 1, urban areas show more responsive activity than mountainous areas, particularly with greater response indicator numbers around the San Francisco Bay and Los Angeles counties, as opposed to Fresno counties. This highlights a potential regional bias and underscores the need to explore their relationship further, which we will address in the next section.

In response to RQ(2), we employed MGWR analysis to explore the relationships between key city-level characteristics and two indicators: awareness and resilience. Table 1 presents the model fit, significance, and bandwidth for the MGWR models. Both models show moderate to high $R^2$ values (0.73 and 0.28 for awareness and resilience indicator model, respectively), indicating a reasonable goodness of fit. The mean intercepts for awareness (-0.151) and resilience (-0.103) suggest that initial values are slightly negative and require adjustment by explanatory variables. Notably, the $R^2$ for the awareness indicator is substantially higher at 0.73, compared to 0.28 for the resilience indicator. This suggests that selected city-level characteristics explain a larger proportion of the variance in awareness compared to resilience.

Table 1. Local MGWR results

| | MGWR Model | | | | | | | | | |
|---|---|---|---|---|---|---|---|---|---|---|
| | Awareness Indicator (β) | | | | | Resilience indicator (γ) | | | | |
| | mean | min | max | STD | bandwidth | mean | min | max | STD | bandwidth |
| Intercept | -0.151*** | -0.20 | -0.09 | 0.04 | 107 | -0.103 | -0.16 | -0.04 | 0.04 | 107 |
| Racial/ethnic | | | | | | | | | | |
| Black or African American | 0.03 | -0.01 | 0.07 | 0.03 | 107 | 0.005 | 0.02 | 0.06 | 0.02 | 107 |
| Asian | 0.17 | 0.14 | 0.19 | 0.02 | 107 | 0.20 | 0.13 | 0.24 | 0.05 | 107 |
| American Indian and Alaska Native | 0.051* | -0.22 | 1.05 | 0.31 | 58 | -0.065 | -0.13 | 0.27 | 0.08 | 76 |
| Hispanic | 0.114** | -0.01 | 0.28 | 0.12 | 99 | 0.204*** | 0.09 | 0.34 | 0.10 | 105 |
| Socioeconomics | | | | | | | | | | |
| Median Household Income | -0.02 | -0.04 | -0.01 | 0.01 | 107 | -0.05 | -0.10 | 0.01 | 0.04 | 107 |
| Unemployment rate | -0.209* | -0.62 | 0.64 | 0.40 | 54 | -0.237* | -0.75 | 0.31 | 0.36 | 58 |
| Multi-unit House | 0.08 | 0.00 | 0.16 | 0.06 | 101 | 0.09 | 0.01 | 0.20 | 0.08 | 105 |
| Without Vehicle | -0.07 | -0.10 | -0.05 | 0.02 | 107 | -0.13 | -0.18 | -0.10 | 0.03 | 107 |
| Female Headed Home | 0.161* | 0.008 | 0.32 | 0.05 | 68 | 0.189 | 0.13 | 0.23 | 0.03 | 107 |
| Demographics | | | | | | | | | | |
| Population Sex ratio | 0.184** | 0.04 | 0.26 | 0.06 | 95 | 0.202 | -0.12 | 0.26 | 0.07 | 89 |
| Urbanized Population | -0.372* | -0.51 | -0.19 | 0.11 | 58 | -0.004 | -0.03 | 0.03 | 0.02 | 107 |
| Wildfire | | | | | | | | | | |
| Burned area | -0.09 | -0.11 | -0.06 | 0.02 | 107 | -0.14 | -0.17 | -0.11 | 0.03 | 107 |

| | | |
|---|---|---|
| No. of observations | 109 | 109 |
| $R^2$ | 0.73 | 0.28 |
| AIC | 218.09 | 320.37 |
| Note: "***" means that variables are significant at the global (state) scale; "**" means that variables are significant at the sub-state regional scale; and "*" means that variables are significant at the local scale. Significance in this table means less than 5%. | | |

As shown in Table 1, racial/ethnic groups display varying correlations with the two indicators, while the other factor groups exhibit more consistent correlations. For example, socioeconomic factors such as Median Household Income, Unemployment Rate, and Without Vehicle consistently show negative correlations with both indicators, whereas factors like Multi-unit House and Female-Headed Home exhibit positive correlations in both models. In general, cities with higher Black or African American, Asian, and Hispanic populations tend to report more positive experiences in terms of both awareness and resilience. However, areas with predominantly American Indian and Alaska Native (AIAN) populations tend to show lower resilience.

While Table 1 presents average parameter estimates, it does not capture regional variations. For instance, the bandwidths for AIAN and Hispanic variables indicate significant sub-state regional variations (Table 1). Further spatial analysis, as shown in Figure 2 and Figure 3, illustrates how the relationships between city-level characteristics and both indicators vary across regions. In both figures, we focus on factors with stronger correlations to the two indicators, using sequential color scales to represent parameter estimate values. Warm colors (primarily red) indicate significantly positive correlations, cool colors (primarily blue) represent significantly negative correlations, and gray areas denote insignificant parameters or missing data.

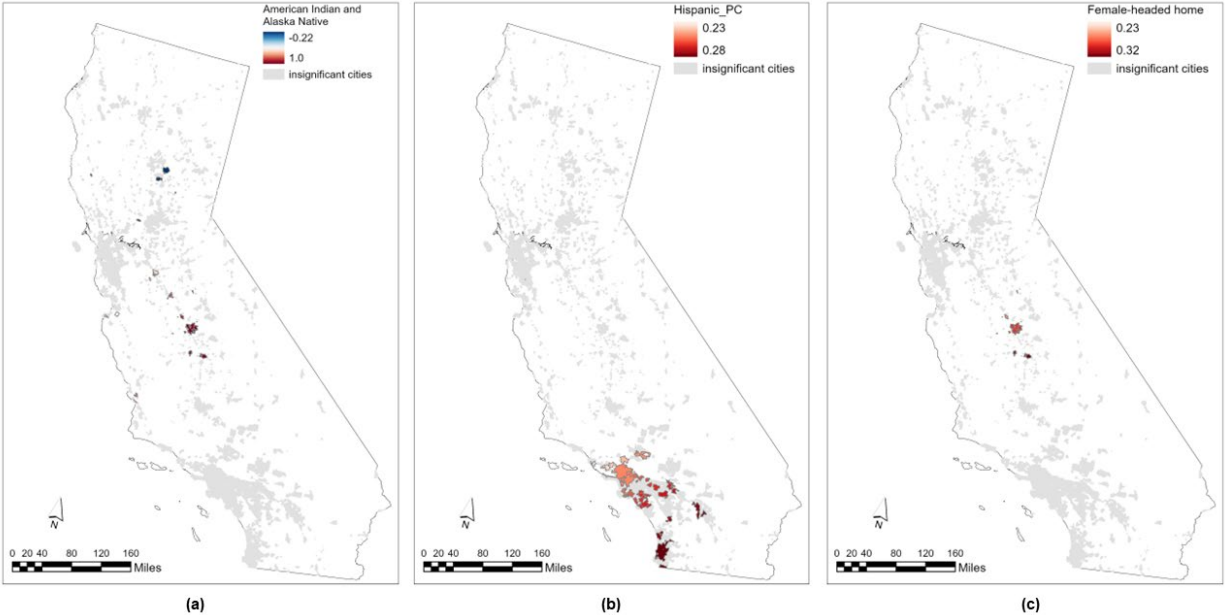

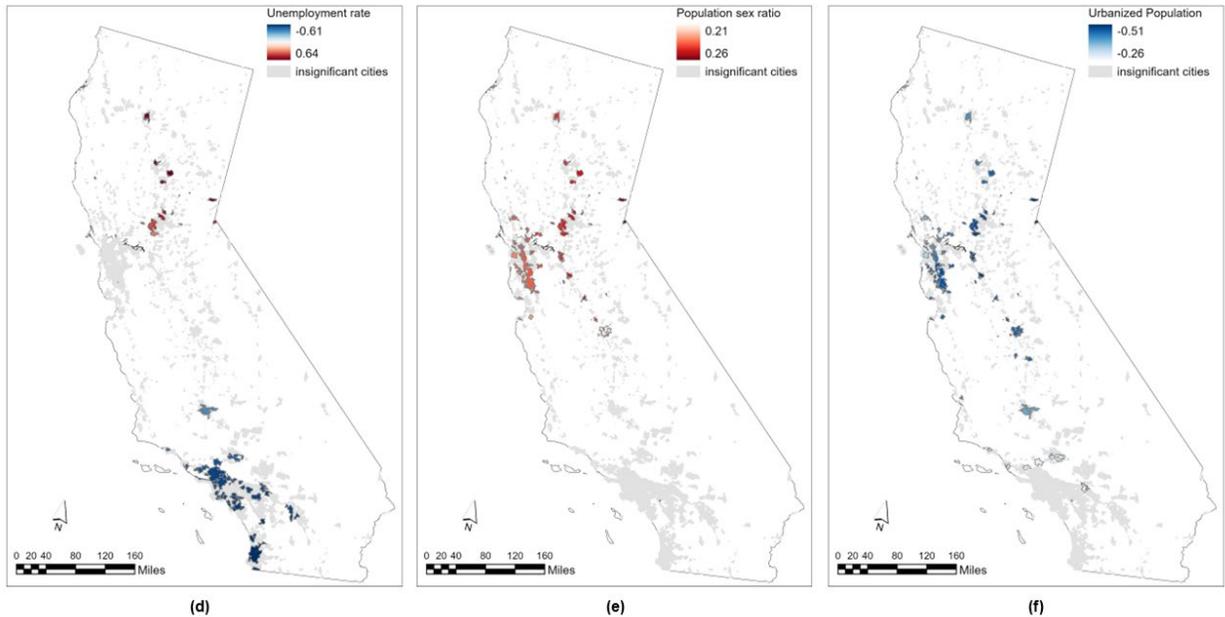

Figure 2. Relationships between independent variables and wildfire awareness indicator at the city level

Figure 2 represents six city-level characteristics that show strong correlations with public awareness indicators. Specifically, minority ethnic groups exhibit positive parameter estimates, with AIAN and Hispanic populations showing significant effects. For AIAN populations, positive correlations are concentrated in central regions, such as Fresno County, whereas negative correlations appear in Northern California cities like Berry Creek and Palermo. Meanwhile, cities near Los Angeles and San Diego reveal that larger Hispanic populations are associated with higher public awareness of wildfires.

Among the socioeconomic factors, Central California cities demonstrate that higher percentages of female-headed households are positively correlated with awareness indicators (Figure 2(c)), suggesting greater awareness in these households. A mixed pattern emerges in the relationship between unemployment rates and awareness, with Southern California showing negative correlations, while Northern California displays positive correlations (Figure 2(d)), highlighting the challenges of raising wildfire awareness in the southern part of the state.

Demographic characteristics reveal that the population sex ratio (males per female) correlates differently with wildfire awareness indicators across different regions of the state. In northern and central areas, cities with more males tend to display greater wildfire awareness (Figure 2(e)). Conversely, urbanized areas such as San Francisco, Fresno, and Sacramento show a negative correlation between the percentage of the urbanized population and wildfire awareness Figure 2(f)). It is interesting to note that there is no significant correlation between wildfire itself and public responses.

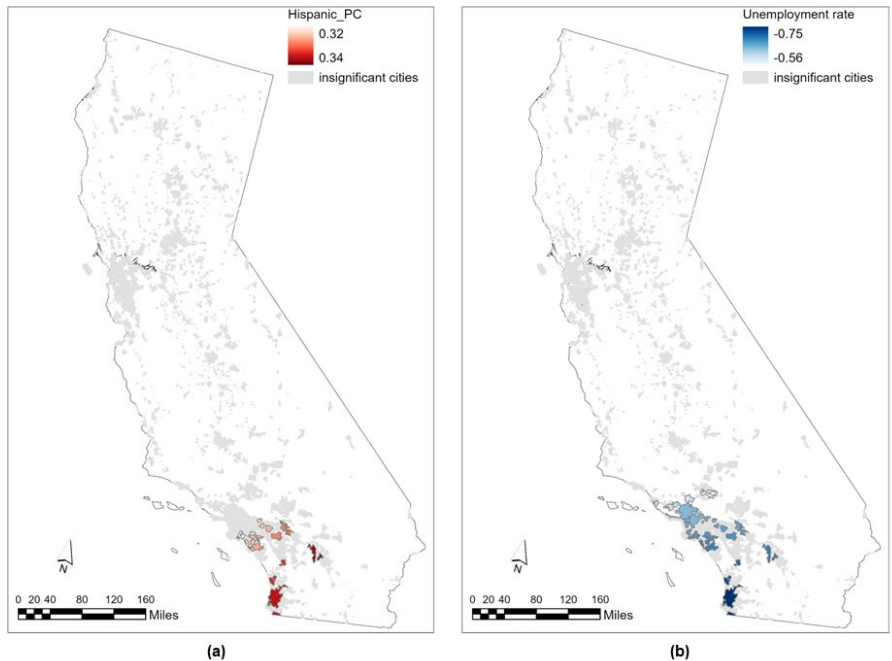

Figure 3. Relationships between independent variables and wildfire resilience indicators at the city level.

In terms of the resilience indicator, Figure 3 highlights two significant factors related to race/ethnicity and employment status that contribute to public resilience. Cities with larger Hispanic populations, particularly in Southern California, show heightened resilience, as indicated by the areas marked in red (Figure 3(a)). This suggests a stronger sense of community resilience, reinforcing the earlier finding of a positive correlation between Hispanic populations and wildfire awareness. Additionally, a negative correlation is observed between unemployment rates and resilience (Figure 3(b)). This pattern becomes more pronounced further south, where higher unemployment rates correspond to lower levels of resilience in response to wildfire.

**Discussion**

1. The insights of social media in wildfire response measurements

Our study integrated compartmental models with social media data to develop two innovative indicators of wildfire response. These indicators, closely aligned with wildfire perimeters, underscore social media's critical role in the real-time monitoring of public reactions, particularly amid the dynamic challenges of WUI fires. This alignment allows decision-makers to rapidly tailor their strategies according to the severity of the wildfire impacts, ensuring that communications and interventions are both timely and effective.

Furthermore, we adopted a ratio-based methodology to measure user engagement, greatly enhancing our ability to detect and address geographical variations in community response levels. While most study areas showed moderate to high response levels, our analysis uncovered significant disparities in response patterns across different regions. Such variations are crucial for decision-makers to identify areas needing urgent support, enabling targeted and efficient deployment of emergency services. As WUI cities expand, our model's adaptability to new data proves vital. By understanding community reactions and communication during these events, decision-makers can better allocate resources to proactively mitigate wildfire impacts. Our approach ensures wildfire management strategies and policies are grounded in community response patterns, promoting more effective and sustainable outcomes.

Additionally, our findings suggest that high-aware regions typically exhibit strong resilience. This correlation indicates that well-informed communities are better prepared and more capable of proactive risk mitigation, likely due to established evacuation plans and a solid understanding of defensible space. Thus, it is essential for decision-makers to enhance public awareness and foster participation in response efforts like volunteer firefighting and information sharing. Such collective actions not only strengthen community resilience but also expedite recovery following disasters. Regions displaying high resilience also offer valuable insights that can be applied to fortify other communities and tackle broader climate change issues.

2. The insights of socio-economic-demographic associations in WUI cites

Our study utilized global and local regression models to examine the city-level characteristics associated with our proposed wildfire response indicators. The local MGWR models, which showed better goodness of fit, revealed significant spatial heterogeneity, highlighting the need for community-specific wildfire management strategies. The in-depth analysis of awareness and resilience indicators further captured important spatial variations and mixed patterns.

First, the findings indicated higher awareness levels among cities with larger racial/ethnic minority populations, particularly AIAN populations in Central California and Hispanic populations in Southern California. This heightened awareness may be attributed to several factors, for example, proximity to wildfire-prone areas in these regions (Yadav et al. 2023) may drive increased outreach efforts, educational initiatives, or culturally-informed safety practices (Collins and Bolin 2009). However, we observed lower awareness in Northern WUI cities with higher AIAN populations, highlighting ongoing challenges in understanding and responding to wildfire measures (e.g., evacuation warnings), which are often issued only in English. Moreover, wildfire preparedness materials and social media content are frequently unavailable in their preferred languages (Méndez, Flores-Haro, and Zucker 2020; Yadav et al. 2023). Therefore, decision-makers must prioritize language accessibility and culturally sensitive outreach to improve preparedness and response in these communities.

Second, socioeconomic vulnerability indicators exhibit a complex relationship with wildfire awareness. Specifically, Southern California cities, such as Los Angeles and San Diego, with higher unemployment rates, exhibited lower awareness. This may be attributed to their prioritization of immediate economic needs over long-term risk awareness, which limits their access to information and resources (Sun et al. 2024; Teo et al. 2018). These findings underscore the need for targeted outreach and education to improve preparedness in economically disadvantaged communities. In contrast, Northern California's positive correlation may result from agricultural and forestry networks keeping unemployed populations informed. Decision-makers should use community workshops in the south and leverage agricultural networks in the north. In Central California, female-headed households emerged as significant contributors to community responses, emphasizing their role in promoting safety practices and information dissemination.

These observations emphasize that socioeconomic differences must be taken into account when developing regional wildfire mitigation strategies.

Further analysis showed that demographic factors, including urbanization and gender dynamics, significantly influenced wildfire awareness differently across regions. For instance, higher sex ratios (more males per females) in Northern and Central California were positively correlated with wildfire awareness, suggesting potential gender disparities in risk perception and crisis communication (Seager 2014). Notably, more urbanized cities such as San Francisco showed lower wildfire awareness, potentially due to a perceived sense of safety from urban infrastructure or limited direct exposure to wildfires (Mockrin, Fishler, and Stewart 2020). However, as wildfires increasingly impact urban areas (Brown and Tousey 2021; Clark, Nkonya, and Galford 2022), outreach efforts must prioritize urban residents to correct misconceptions and enhance awareness of wildfire risks.

Fourth, our analysis found no significant correlation between wildfire intensity attributes and public responses, regardless of awareness and resilience. Although previous research suggests that wildfire risk is often linked to community vulnerability and disproportionately affects different communities (Hwang and Meier 2022; Mahmoud and Chulahwat 2018), our findings emphasize the complexity of these relationships. The absence of a direct correlation suggests that more dynamic, context-specific factors may influence public response. Therefore, decision-makers should consider a case-by-case approach to wildfire risk assessment and response planning, rather than relying solely on generalized wildfire intensity metrics.

Last, our findings suggest that certain minority groups and low-unemployment areas are more adaptive to wildfires. For example, Southern California cities with higher Hispanic populations show positive resilience, aligning with previous research suggesting that ethnic minority groups near Coastal Sage socio-ecoregion are often more attuned to environmental issues (Lazri and Konisky 2019; Yadav et al. 2023). Cities with higher unemployment rates show a negative relationship with resilience, likely due to limited resources among the unemployed, reducing their ability to recover from wildfire impacts (Lambrou et al. 2023). These results highlight ongoing issues of environmental justice, where disadvantaged groups face higher risks and reduced resilience, exacerbating existing socioeconomic inequities. Addressing these disparities calls for targeted interventions focused on resource allocation and support systems to bolster resilience among vulnerable WUI communities.

3. Limitations and future work

The current study has several limitations and potential areas for future research. First, Twitter data may not fully represent the general population due to biases in demographic biases, as the platform's user base tends to skew younger, more educated, and predominantly White and male (Mitchell 2021). These biases may limit the accuracy and generalizability of our findings. Future research should explore strategies to enhance data representativeness, such as incorporating data from multiple social media platforms, applying demographic weighting adjustments based on Census data, or supplementing social media data with targeted surveys of underrepresented groups.

Next, we utilized registration and content-based locations to assess disaster responses, which may not accurately reflect users' locations. Users might post tweets that imply locations unrelated to their registered addresses. Additionally, relying on Named Entity Recognition (NER) for identifying content-based locations can lead to errors in location identification. Future research could explore the use of more advanced tools, such as large language models (LLMs), to improve location information extraction.

Additionally, applying the SIR model to social media data has challenges. We excluded cities with sporadic or minimal tweeting activity, which could overlook insights into the wildfire response dynamics in these areas. To address this, future studies could integrate data from alternative platforms to enrich the dataset or employ methodologies to handle missing data effectively. This approach would allow for a more comprehensive analysis and potentially uncover valuable patterns and trends in wildfire response across a broader spectrum of communities.

Furthermore, our wildfire response assessment was conducted at the city level, and due to limitations in the location information obtained from social media, we could not perform a more granular geospatial analysis. As a result, there may be some challenges in aligning city boundaries with finer geographic units, such as tract or block. Moreover, aggregating independent variables can obscure finer spatial variations and mask underrepresented areas with concentrated risks or disparities due to the non-uniform distribution of demographic, socioeconomic, and exposure characteristics. Future research should focus on collecting more precise and consistent media or spatial data at similar geographic units to improve accuracy. Exploring alternative methods for integrating data across mismatched geographic boundaries could also enhance our understanding of the complexities involved in wildfire response and risk.

Lastly, while the MGWR model was utilized to capture the spatial heterogeneity in relationships between response variables and socioeconomic factors, it has certain limitations. One challenge is that interpreting MGWR results can be more complex compared to simpler spatial models, requiring careful attention to how relationships vary across space. Although MGWR allows for different spatial scales by relaxing the assumption of global stationarity, it still assumes local stationarity within regions. This means that it may not fully capture the variability and more subtle differences that exist across different locations. Future research could focus on refining spatial models to better account for these local variations and improve the accuracy of regional analyses.

**Materials and Methods**

Figure 4 outlines the research framework employed in this study. To address the first research question, we collected data from Twitter, employing natural language processing and topic modeling to identify information relevant to the responses during the 2020 California wildfires. These topics were integrated with compartmental epidemiological models to estimate public response levels for each community. From this integration, we defined two indicators for public awareness and resilience; these indicators have been verified in past studies (Ma, Li, Hemphill, et al. 2024). Meanwhile, we compiled sociodemographic factors and wildfire characteristics at the city level to address our second research question. Utilizing the public response indices derived from the SIR model, we applied Multiscale Geographically Weighted Regression (MGWR) model to explore the relationship between these factors and public responses, which allowed us to account for spatial heterogeneity in the data and provided a more localized understanding of public reactions.

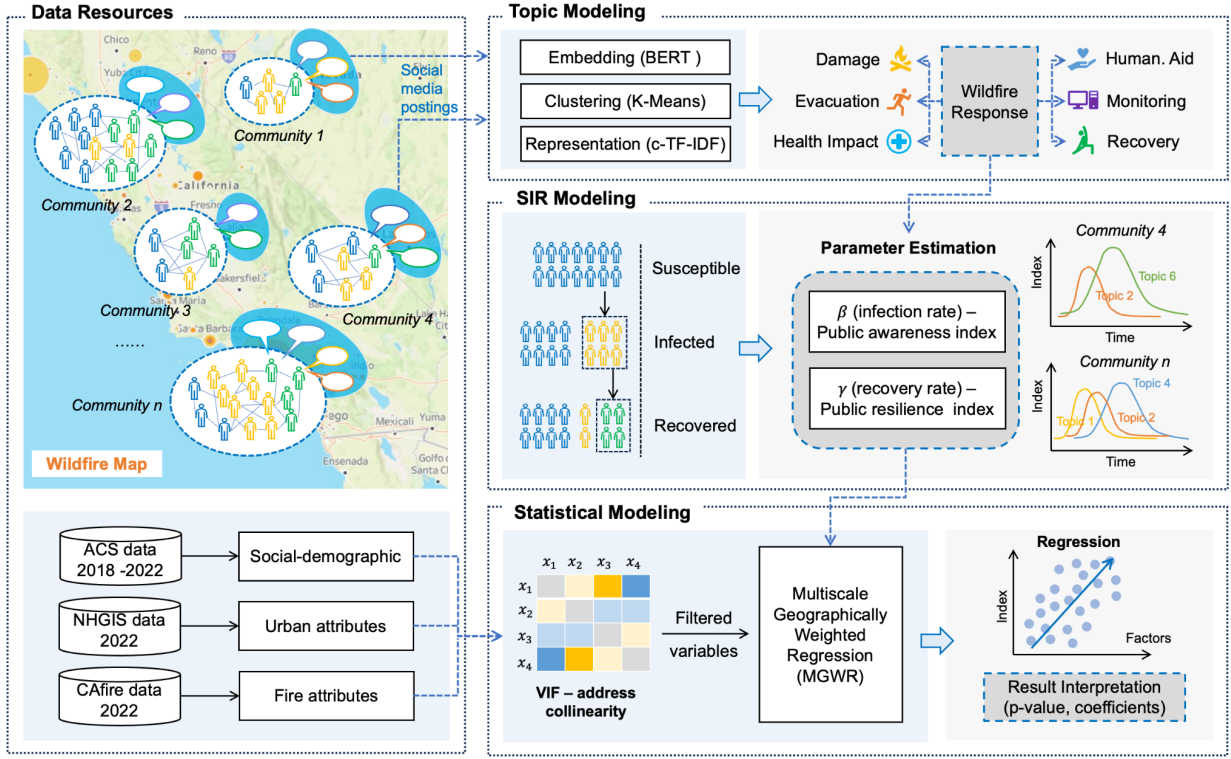

Figure 4. The research framework

1. Data collection and preprocessing

This research utilized two primary types of datasets. The first consisted of social media data from Twitter. Compared to other major American platforms, Twitter (now X) stands out as a source for breaking news, with 75% of users reporting that they obtain real-time information there, compared to smaller shares of users on Facebook (58%), TikTok (55%), and Instagram (44%) (Matsa 2024b). Moreover, three-quarters or more of Twitter users (85%) say they see opinions on current events (Matsa 2024a), making Twitter a valuable resource for capturing public reactions in real-time. As such, we employed the Twitter API to collect tweets related to "wildfire" in English from September 2 to October 4, 2020, resulting in 932,325 records, including 175,946 original tweets across the entire U.S. While we acknowledge that Twitter users communicate in various languages, the majority of users in the U.S. (approximately 80%) use English as their primary language (Antonakaki, Fragopoulou, and Ioannidis 2021; Midha 2014). Therefore, to simplify our analysis and ensure consistency in the interpretation of data, we primarily focused on English-language content.

Twitter typically provides three location types: 1) content-based location, 2) posting location, and 3) registration location (i.e., home location). For our analysis focusing on CA state, we primarily used registration locations provided in user profiles, which previous studies suggest can effectively proxy event locations (Ao, Zhang, and Cao 2014; Ozdikis, Oguztuzun, and Karagoz 2013; Rachunok et al. 2022). Our dataset included 37,123 tweets including registration location at city level. To supplement this, we also utilized content-based locations, which prior studies have shown to correlate well with users' actual locations (Paradkar et al. 2022). We employed the SpaCy Named-Entity Recognition (NER) tool to extract content-based locations, identifying 25,124 tweets containing such city names. In cases where tweets included both content and registration locations, we retained only the registration location information. This resulted in a total of 59,100 tweets, with 19,740 original tweets located in CA state for further analysis.

The second dataset includes city-level characteristics. Based on prior research emphasizing racial/ethnic, socioeconomic, and demographic characteristics (Table A1), we selected 18 key variables from the American Community Survey (ACS) 2020 profile tables (2016–2020 estimates) (US Census Bureau 2024) the National Historical Geographic Information System (NHGIS) [https://www.nhgis.org/]. These variables, such as Black/African Population percentage, median household income, unemployment rate, population aged 65+, percentage with a Bachelor or higher, have been categorized into three groups, including racial/ethnic, socioeconomic, and demographic.

To align the location information with tweet data, we utilized Tigris package in R (Walker and Rudis 2024) to assign GEOIDs to cities based on the 2020 Census. While ACS provides variables at smaller units like block groups, city-level estimates are generally more reliable due to smaller margins of error (Spielman, Folch, and Nagle 2014). We also aggregated tract-level urbanized population data from NHGIS with spatially overlapping city boundaries for more accurate city-level estimates.

Additionally, we considered two wildfire intensity attributes: wildfire occurrence and incident size, as wildfire activity itself may influence public responses. We retrieved 2020 wildfire data from CaFire [https://www.fire.ca.gov/], which contains geolocation data (latitude and longitude). To ensure consistency in location information, we employed ArcGIS Pro to aggregate the burned acres of wildfires to city level and calculated the burned percentage of each city. A dummy variable was introduced to represent wildfire occurrence, coded as 1 if wildfires were recorded and 0 if none occurred. These two factors were grouped under the "Wildfire" category, bringing the total number of variables to 20.

2. Response indicators

To develop numerical metrics that capture community responses to wildfires and to examine their geographic distribution, we first need to identify tweets related to wildfire responses. Prior research (Kabra, Ghosh, and Mukherjee n.d.; Li, Haunert, et al. 2023; Ma, Li, Mao, et al. 2024; Rachel J et al. 2024; Zhou et al. 2023) has demonstrated that topic modeling is an effective method for uncovering patterns in public reactions to crises. Unlike supervised machine learning, which requires extensive manual annotation, topic modeling provides a faster, automated approach for tracking and categorizing responses throughout the disaster lifecycle. Traditional methods like LDA rely on probabilistic models that assume each document is a mix of latent topics, with each word assigned to one. However, LDA's dependence on statistical inference can lead to less accuracy when applied to short texts in social media.

Table 2. Descriptive Statistics of the Variables

| Variable | | Description | Mean | St.d | Min. | 50% | Max. |
|---|---|---|---|---|---|---|---|
| **Dependent Variables** | | | | | | | |
| **Infection rate** | Awareness Indicator ($\beta$) | The rate of susceptible populations engages in wildfire responses | 1.28 | 1.50 | 0.00 | 1.01 | 9.35 |
| **Recovery rate** | Resilience Indicator ($\gamma$) | The rate of infected populations disengages in wildfire responses | 0.73 | 0.63 | 0.00 | 0.70 | 3.18 |
| **Independent Variables** | | | | | | | |
| **Racial/ethnic** | Black/African American (%) | The percentage of the Black or African American population | 4.26 | 4.55 | 0.00 | 2.50 | 22.70 |
| | Asian (%) | The percentage of the Asian population | 14.35 | 13.79 | 0.00 | 17.50 | 69.20 |
| | American Indian and Alaska Native (%) | The percentage of American Indian and Alaska Native population | 0.70 | 0.72 | 0.00 | 0.60 | 5.60 |
| | Hispanic (%) | The percentage of Hispanic or Latino population | 28.5 | 19.30 | 0.00 | 23.60 | 97.5 |
| **Socioeconomics** | Median Household Income ($ in 10k) | The median household income (in 2020 Inflation-Adjusted Dollars) | 8.948 | 3.331 | 2.891 | 8.411 | 18.903 |
| | High Educated (%) | Percentage of the population aged 25 and over with a bachelor's degree or higher | 42.31 | 19.25 | 2.40 | 41.60 | 93.90 |
| | Unemployment rate (%) | Percentage of the civilian labor force that was unemployed | 5.68 | 2.69 | 0.00 | 5.20 | 22.50 |
| | Mobile Home (%) | Percentage of total housing units that are mobile homes | 3.57 | 6.28 | 0.00 | 2.10 | 43.80 |
| | Without Vehicle (%) | Percentage of occupied housing units with no available vehicles | 6.45 | 4.30 | 0.00 | 5.70 | 30.80 |
| | Crowd Home (%) | Percentage of total housing units in structures of more than 10 units | 16.49 | 10.47 | 0.00 | 14.60 | 54.10 |
| | Female Headed Home (%) | Percentage of total households with a female householder and no spouse or partner present | 27.26 | 5.61 | 12.10 | 26.80 | 53.60 |
| | GINI index | A summary measure of income inequality, ranging from 0(perfect equality) to1(perfect inequality) | 0.46 | 0.05 | 0.20 | 0.46 | 0.62 |
| **Demographics** | Age under 18 (%) | Percentage of residents under 18 years | 20.86 | 5.92 | 0 | 21.50 | 35.4 |
| | Age over 65 (%) | Percentage of residents over 65 years | 16.64 | 6.99 | 3.90 | 15.10 | 45.30 |
| | Population Sex ratio | Sex ratio (number of males per 1 female) | 0.9805 | 0.1428 | 0.767 | 0.964 | 2.114 |
| | Population Density (plp/$m^2$) | The number of people per square meter | 0.0017 | 0.0011 | 0.00 | 0.00 | 0.01 |
| | Population with a disability (%) | The percentage of the disability population | 10.78 | 4.40 | 0 | 10.1 | 30.6 |
| | Urbanized Population (%) | Percentage of the population lived in urban areas | 92.00 | 21.75 | 0.00 | 98.84 | 100.00 |
| **Wildfire** | Occurrences | Whether the wildfire was recorded or not at a city | 0.25 | 0.43 | 0.00 | 0.00 | 1.00 |
| | Burned area (%) | The percentage of area burned by wildfires | 0.66 | 2.09 | 0.00 | 0.00 | 14.43 |

To address this limitation, we adopted the methodology outlined by (Ma, Li, Hemphill, et al. 2024) employing a transformer-based topic modeling approach known as BERTopic (Grootendorst 2022) to identify tweets relevant to wildfire responses. Unlike LDA, BERT-based models can capture words' semantic relationships and contextual meanings. Following the default embedding tool (i.e., Sentence-BERT (SBERT)) and dimension reduction tool (i.e., Uniform Manifold Approximation and Projection (UMAP)), we adopted k-means algorithms for our clustering method. We determined the optimal number of clusters for topic modeling using K-means clustering with the elbow method. The elbow method is a visual technique that helps identify the optimal number of clusters (K) by checking the Within-Cluster Sum of Squares (WCSS), which measures how close points are to their cluster centers by summing the squared distances. To find the optimal number of clusters, we tested different values of K ranging from 2 to 200. Overall, we found the WCSS significantly dropped at the number of 50 clusters (Figure A2). Therefore, we determined that 50 clusters represented an optimal balance between clustering granularity and topic interpretability. The 50 topics were further categorized into a smaller number of topics to ensure a more streamlined analysis and easier interpretation of the data. Subsequently, two authors manually reviewed the machine-generated representative keywords, which were derived utilizing class-TF-IDF within each cluster, and assigned topics to parent categories. Upon reaching a consensus on each topic and drawing support from a literature review (Table A3), we identified six wildfire response-related topics. Details of these topics are listed in Table A2. All tweets categorized under these topics were thus considered to pertain to wildfire responses.

After that, we applied the SIR epidemiological model to measure public engagement as a proxy for public response. We observed similarities between information dissemination and disease transmission, where involvement in a topic (akin to infection by a virus) depends on initial interest (contact with the virus). The SIR model classifies the population into three phases: (1) Susceptible, (2) Infective, and (3) Recovered. In social media terms, users who might engage in a topic are deemed "susceptible," indicating a potential willingness to partake in discussions by posting tweets. Those who actively engage in discussions are categorized as "infected." Over time, users may lose interest or cease participation, transitioning into the "recovered" phase.

Using this model, we introduced two key indicators: public awareness and resilience. Evidently, higher public engagement via social media significantly enhances situational awareness—the collective understanding of an event's potential impact (Kankanamge, Yigitcanlar, and Goonetilleke 2020; Liang and Ramirez-Marquez 2024). During disasters, situational awareness is crucial, as well-informed communities are more confident and better equipped to respond effectively. In our SIR model, the transition from "susceptible" to "infected," (infection rate ($\beta$)) indicates public awareness level—capturing how quickly users engage with wildfire-related content through activities like following, sharing, or posting. A higher infection rate means more rapid public involvement and heightened awareness.

Resilience, on the other hand, reflects the community's ability to cope, adapt, and return to stability following a disaster (Xie, Pinto, and Zhong 2022). The transition from "infected" to "recovered," quantified as the recovery rate ($\gamma$), serves as an indicator of resilience. This metric indicates how users disengage from discussions, often signifying issue resolution, information saturation, or a shift in attention. These dynamics are mathematically modeled as follows:

$$S(t) = \frac{dS}{dt} = -\frac{\beta IS}{N}$$

$$I(t) = \frac{dI}{dt} = \frac{\beta IS}{N} - \gamma I$$
$$R(t) = \frac{dR}{dt} = \gamma I$$

Here, $S(t)$, $I(t)$, and $R(t)$ represent the number of susceptible, infected, and recovered users at the time $t$, respectively, $N$ denoting the total user base engaged in the topic, with $\beta$ as the awareness indicator, and $\gamma$ as the resilience indicator.

3. Regression Analysis

To examine the relationships between public wildfire responses, wildfire risks, and socio-economic-demographic characteristics, we developed two regression models: the global Ordinary Least Squares (OLS) model and the local Multiscale Geographically Weighted Regression (MGWR). In our models, we treated two response indicators—awareness and resilience—as dependent variables, and city-level characteristics as independent variables. These variables and their descriptions are summarized in Table 2.

The OLS model employs the least squares estimation method for parameter estimation and is traditionally non-spatial, meaning it does not account for spatial variations in the data (Wooldridge 2015). The general equation for the OLS model is:

$$y_i = \beta_0 + \beta x_i + \varepsilon_i,$$

*where*, $y_i$ represent the dependent variables ((awareness and resilience indicators) in city i). $\beta_0$ is the intercept; $x_i$ denotes the vector of selected explanatory variables; $\beta$ is the vector of regression coefficients; and $\varepsilon$ is a random error term.

MGWR is an advanced form of Geographically Weighted Regression (GWR) that allows each relationship to occur at its own spatial scale, effectively addressing issues of scale misspecification common in GWR (Fotheringham, Yang, and Kang 2017). This model is structured as a generalized additive model where each variable can have a distinct bandwidth, reflecting the local context for each regression analysis. The MGWR model's equation is:

$$y_i = \sum_j \beta_{ij}(u_i, v_i) X_{ij} + \varepsilon_i, i = 1, 2, \cdots, N,$$

*where*, $y_i$ are the dependent variables, namely awareness and resilience indicators in city i. $\beta_{ij}(u_i, v_i)$ represents the value of the j-th regression parameter, which varies across the feature space. $X{ij}$ is the j-th explanatory variable at location i. $\varepsilon$ is a random error term.

The calibration of the MGWR model produces a vector of optimal bandwidths for each variable, helping to describe the spatial scales at which different processes vary. The global scale suggests a bandwidth large enough to include most cities, providing consistency across the study area. Regional and local scales indicate the inclusion of fewer cities, reflecting spatial heterogeneity.

We addressed potential multicollinearity issues by removing highly correlated variables and further calculated the Variance Inflation Factor (VIF) values to confirm the absence of multicollinearity (O'brien 2007). Further details on this testing are provided in Figure A5. Following this rigorous selection process, we identified a set of 12 explanatory variables with VIF values smaller than 5. These variables include Black or African American, Asian, American Indian and Alaska Native, Hispanic, Median Household Income, Unemployment rate, Multi-unit House, Without Vehicle, Female-Headed Home, Population Sex ratio, Urbanized Population, and Burned area.

Model performance was compared using $R^2$ and Akaike Information Criterion (AIC) values, with local models generally showing higher $R^2$ values and lower AIC values (Table A5 and Table 2), indicating better performance and explanatory power. The spatial clustering of public awareness and resilience indicators, discussed in Figure A6, favored local models due to their sensitivity to spatial characteristics. Overall, the MGWR model demonstrated superior performance due to its ability to accommodate multiple scales and accurately reflect local variations, making it the preferred model for further analysis in our study.

# Appendix

## Appendix A1

**Table A1 shows the literature with corresponding variables that support our city-level characteristics selection.**

Table A1: Literature supports for city-level characteristics selection

| Author and Year | Disaster/Hazard type | Independent variables |
|---|---|---|
| Chakraborty et al. (2023) | Flood | Race and ethnicity (such as Latin American, Asian individuals, and American Indian or Alaska Native (AIAN) individuals); Socio-economic status (such as population with a college degree, low-income status, unemployment, number of rooms in housing, estimated home value); demographics (such as older adults, gender, and individuals with disabilities). |
| Lambrou et al. (2023) | Wildfire | Race and ethnicity (i.e. White, Black, Native, Asia, Hispanic individuals); Socio-economic status (such as female-headed households, education, mobile homes, multi-unit housing, lack of vehicle ownership, overcrowding in housing units, home value, poverty, income, and employment); demographics (such as language, citizenship, disability status, gender, and age). |
| Hu, Feng, and Sun (2023) | Toxic Chemical Hazard | Race and ethnicity (such as Black, Asian, and Hispanic individuals; Socio-economic status (such as median household income, adults without a high school diploma, unemployment, and Gini index); |
| Yadav et al. (2023) | Wildfire | Socio-economic status (such as annual income, poverty, and population with a high school degree); Demographics (such as older adults, disabled population, and language use). |
| Forati and Ghose (2022) | Hurricane | Race/ethnicity (i.e. White, Black, Native, Asia, Hispanic individuals); Socio-economic status (such as median income); Demographics (such as median age) |
| Teo et al. (2018) | General disaster (including tropical cyclones, severe weather events, riverine flooding, coastal inundation, heatwaves, bushfires and earthquakes) | Socio-economic status (such as household income and occupations); Demographics (such as gender, household size, language use, and duration of residence). |
| Cutter and Finch (2008) | General disaster | Race/ethnicity (such as Asian, African American, Native American, and Hispanic); Socioeconomic status (such as poverty, population with less than a high school education, per capita income, median house value, female-headed households, and employment rates); Demographics (such as median age, population under 18, population over 65, female, and population density). |

**Appendix A2**

**The elbow method plot (Figure A2) suggests that the optimal number of clusters (K) is around 50. This is where the "elbow" of the curve appears, indicating a significant drop in the sum of squared distances, after which the decrease becomes more gradual. Choosing K around this point provides a balance between clustering granularity and interpretability in the topic modeling process.**

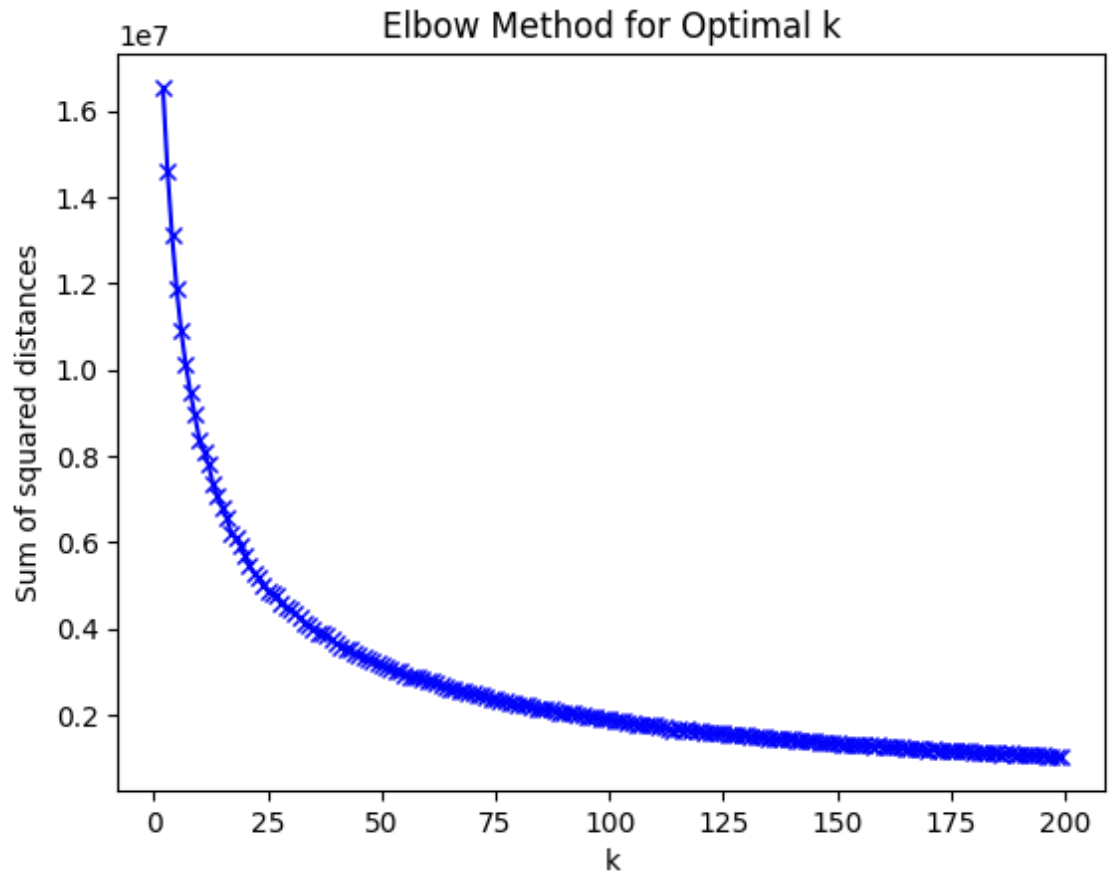

Figure A2: Elbow Method Plot Suggesting Optimal Number of Clusters (K) for Topic Modeling

**Table A2 presents our final topic explanations, their corresponding original machine-generated topic numbers and keywords, along with tweet examples.**

Table A2. Topic identification and examples of representative tweets and keywords.

| Cluster number | Representative keywords | Representative situation | Representative tweet example | Topic |
| --- | --- | --- | --- | --- |
| 0, 1, 5, 9, 10, 12, 16, 21, 37 | smoke, health, lung, air, ash, sky, quality, unhealthy, orange, wind | Health risks, impacts on air quality | *These are not clouds. Wildfire smoke *cough** | Health impact |
| 4, 11, 15, 18, 20, 24, 27, 28, 29, 30, 35, 38, 48 | acre, burn, glassfire, death, kill, school, close, highway, missing, scorch | Infrastructure disruptions, injuries, fatalities, and other consequential social and economic damages | *Wildfire. right lane blocked in #Sherwood on 99w SB near SW Chapman Rd/SW Brookman Rd #PDXtraffic* | Damage |
| 31, 46, 47 | expect, latest, ready, prepare, devastation, real, map, live, time, announce | Live report, real-time map, and other tracking tools and methods | *California, Oregon, and Washington live wildfire maps are tracking the devastation in real time* | Monitoring |
| 2, 7, 22, 23, 25, 32, 43 | evacuate, firefighter, gov, rescue, level, order, battle, response, animal, governor | Evacuation plans, rescue efforts, and other response measures facilitating evacuation. | *All of Clackamas County is under Level 1, 2 or 3 evacuation order* | Evacuation |
| 8 | insurance, housing, risk, family, earthquake, disaster, safe, emergency, plan, homeowner | Infrastructure repair, emotion restoration, community rebuilding, and other recovery efforts | *RT @CAL_FIRE: Returning home after a wildfire can be difficult and the amount of damage is often unknown. Take the time to inspect the outside and inside of your home for any remaining dangers. Learn what to look for by visiting* | Recovery |
| 6 | relief, fund, donation, support, donate, effort, evacuee, victim, recovery, community | Donations, relief fund, and other humanitarian endeavors | *Please give if you can. Wildfire Relief Fund* | Humanitarian aid |
| 3, 13, 36 | party, pyrotechnic, spark, cause, climate, change, forest, management, tree, fuel | User report information about the causes of wildfires | *A couple's plan to reveal their baby's gender went up not in blue or pink smoke but in flames when the device they used sparked a wildfire that burned thousands of acres and forced people to flee from a city east of Los Angeles* | Wildfire causes* |
| 14, 26, 33, 42 | covid, pandemic, spread, virus, protest, resume, vaccine, live, arrested, china | User report information related to COVID-19 pandemic | *#Portland protests resume after wildfire hiatus with Ginsburg vigil, more vandalism* | COVID-19 pandemic* |
| 19 | antifa, arson, rumor, conspiracy, arrested, police, spread, blm, false, theory | User report information that related to the credibility of wildfire message | *Antifa setting wildfires is a conspiracy? See below links: ARSON-ANTIFA Conspiracy. Heh. Pt1* | Misinformation* |
| 17, 34, 39, 40, 41, 44, 45, 49 | fightwave, berry, madman, canadian, flight, foil, baked, photo, span, arrival | User report information that does not fit any above categories | *Oregon man arrested twice in 12-hour span for starting 'multiple' fires near Portland freeway* | Other* |

Note: *These topics did not reflect disaster response efforts and were not included in the following analysis

**Appendix A3**

**Table A3 shows the literature with corresponding topics that support our topic identification within social media data.**

Table A3. Supported literature with corresponding topics.

| Author and Year | Disaster type | Topic in disaster response |
| --- | --- | --- |
| Zhou et al. (2023) | Hurricane | "Advisory," "Casualty," "Damage," "Relief," "Information source," "Emotion," "Animal" |
| Gründer-Fahrer et al. (2018) | Flood | "Information," "Help," "Alertness/relief," "Relaxation" |
| Karami et al. (2020) | Flood | "Victims," "Damage and Costs," "Drinking Water," "Insurance," "Homelessness," "Road Damage," "Roof Damage," "Bridge Damage," "Flood Report," "Power Lost," "Animal" |
| Wang et al. (2015) | Rainstorm | "Traffic," "Weather," "Disaster Information," "Loss and Influence," "Rescue Information" |
| Fan et al. (2020) | Hurricane | "Infrastructure and utility damages," "Affected and injured individuals," "Rescue, volunteering, or donation effort," "Other relevant information" |
| Karimiziarani et al. (2022) | Hurricane | "Caution," "Damage," "Evacuation," "Injury," "Help," "Sympathy" |
| Jin and Spence (2021) | Hurricane | "Medical support," "Resilience," "Food/water supply," "Forecast/update," "Rescue," "Concerns," "Aid," "Fundraising," "Evacuation," "Relief" |
| Huang et al. (2022) | Typhoon | "Warning," "Damage," "Work and life," "Temperature," "Concern and fear", "Traffic," "Caution and advice," "Gratitude," "Weather" |

**Appendix A4**

**The Variation Inflation Factor (VIF) was calculated for all variables in Table A4, with values less than 5, indicating no significant multicollinearity issues. Additionally, we conducted a pairwise Wilcoxon test (Figure A4) to further confirm that multicollinearity was not present among the explanatory variables at this spatial scale.**

Table A4: VIF results

| Variable | VIF | Variable | VIF |
|---|---|---|---|
| Black or African American | 1.309648 | Multi-unit House | 1.419059 |
| Asian | 1.487198 | Without Vehicle | 1.529838 |
| American Indian and Alaska Native | 1.491807 | Female Headed Home | 1.987441 |
| Hispanic | 1.672939 | Population Sex ratio | 1.403827 |
| Median Household Income | 2.479494 | Urbanized Population | 1.951433 |
| Unemployment rate | 1.953860 | Burned area | 1.048445 |

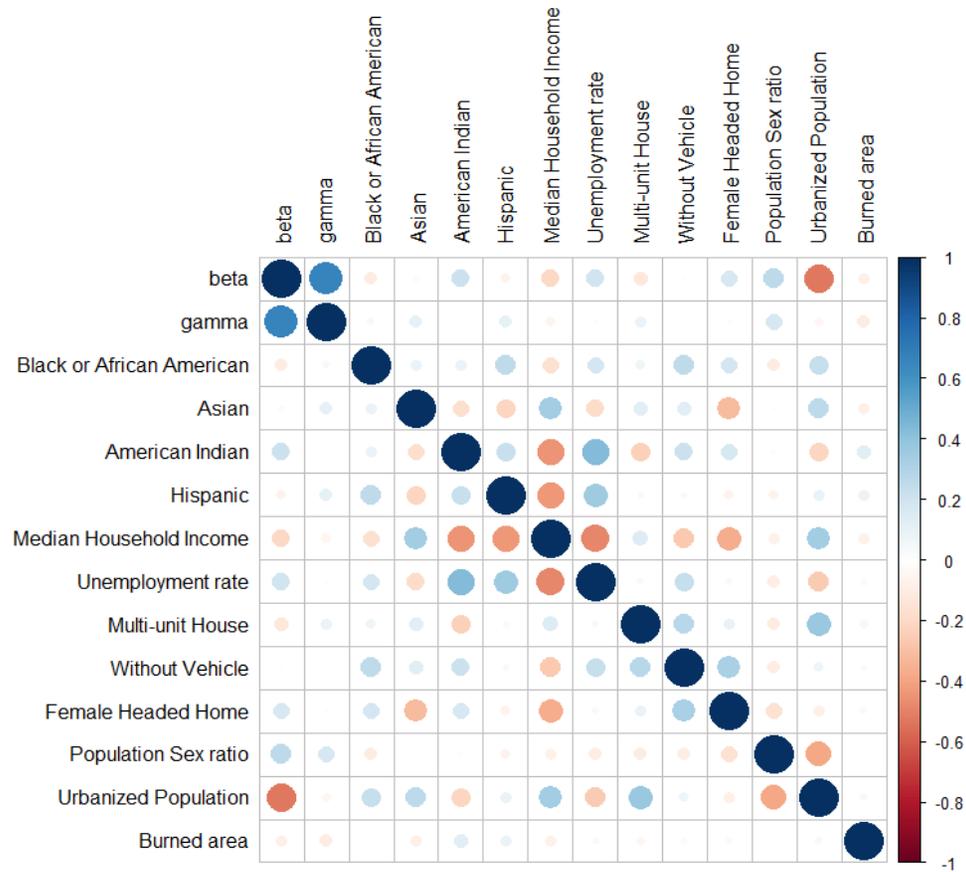

Figure A4: Pairwise correlation results

**Appendix A5**

**Table A5 presents the results of the global OLS models, which show lower R² and Akaike Information Criterion (AIC) values compared to the local models, suggesting reduced model performance and explanatory power. Additionally, Figure A5 displays**

the local Moran's I results, highlighting the spatial clustering of public awareness and resilience indicators. These findings emphasize the importance of accounting for spatial disparities and utilizing local models for more accurate analysis.

Table A5: OLS results

| | OLS model | | | | | |
|---|---|---|---|---|---|---|
| | Awareness indicator ($\beta$) | | | Resilience indicator ($\gamma$) | | |
| | parameter | SE | t-value | parameter | SE | t-value |
| Intercept | 0 | 0.082 | 0 | 0 | 0.097 | 0 |
| Racial/ethnic | | | | | | |
| Black or African American | -0.058 | 0.094 | -0.618 | 0.01 | 0.111 | 0.087 |
| Asian | 0.213** | 0.1 | 2.134 | 0.187 | 0.119 | 1.575 |
| American Indian and Alaska Native | 0.116 | 0.1 | 1.165 | 0.007 | 0.119 | 0.061 |
| Hispanic | 0.018 | 0.106 | 0.171 | 0.154 | 0.126 | 1.221 |
| Socioeconomics | | | | | | |
| Median Household Income | 0.068 | 0.129 | 0.528 | -0.026 | 0.153 | -0.169 |
| Unemployment rate | 0.154 | 0.114 | 1.344 | 0.016 | 0.136 | 0.118 |
| Multi-unit House | 0.06 | 0.097 | 0.611 | 0.105 | 0.116 | 0.904 |
| Without Vehicle | -0.099 | 0.101 | -0.982 | -0.106 | 0.120 | -0.879 |
| Female Headed Home | 0.261** | 0.115 | 2.259 | 0.115 | 0.137 | 0.839 |
| Demographics | | | | | | |
| Population Sex ratio | 0.135 | 0.097 | 1.397 | 0.182 | 0.115 | 1.578 |
| Urbanized Population | -0.461*** | 0.114 | -4.038 | -0.045 | 0.136 | -0.334 |
| Wildfire | | | | | | |
| Burned area | -0.059 | 0.084 | -0.706 | -0.102 | 0.100 | -1.021 |

| | | |
|---|---|---|
| No. of observations | 109 | 109 |
| $R^2$ | 0.092 | 0.357 |
| AIC | 324.864 | 287.128 |
| *Note: \*\*\* Significant at the 1% level. \*\* Significant at the 5% level. \* Significant at the 10% level.* | | |

Figure A5: Local Moran's I results

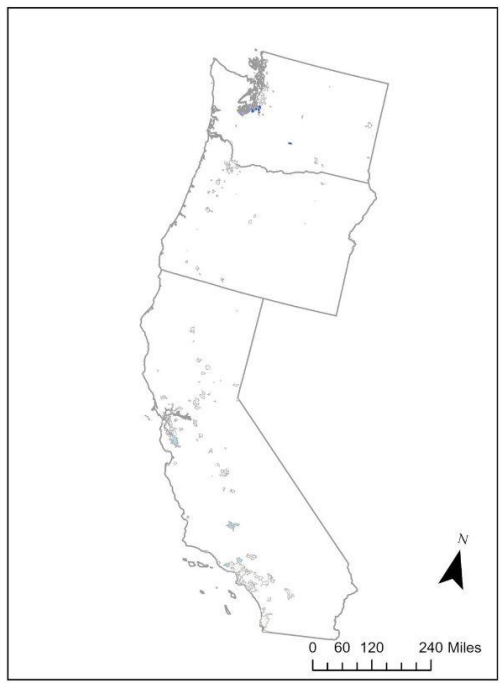
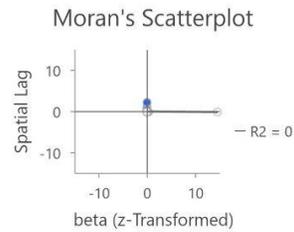
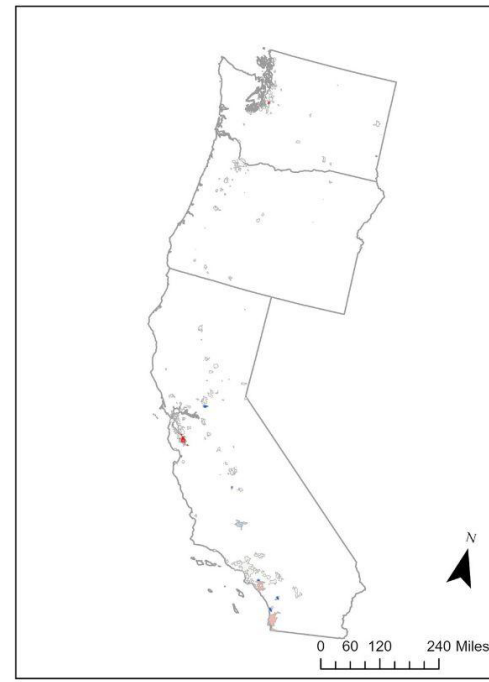
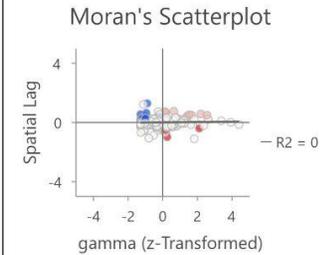